\documentstyle[12pt]{article}
\advance\textheight by \topskip
\begin{document}
\newcommand{\Psl}{\not\!\! P}
\newcommand{\dsl}{\not\! \partial}
\newcommand{\half}{{\textstyle\frac{1}{2}}}
\newcommand{\for}{{\textstyle\frac{1}{4}}}
\newcommand{\eqn}[1]{(\ref{#1})}
\def\la{\mathrel{\mathpalette\fun <}}
\def\a{\alpha}
\def\b{\beta}
\def\g{\gamma}\def\G{\Gamma}
\def\d{\delta}\def\D{\Delta}
\def\e{\epsilon}
\def\et{\eta}
\def\z{\zeta}
\def\t{\theta}\def\T{\Theta}
\def\l{\lambda}\def\L{\Lambda}
\def\m{\mu}
\def\f{\phi}\def\F{\Phi}
\def\n{\nu}
\def\p{\psi}\def\P{\Psi}
\def\r{\rho}
\def\s{\sigma}\def\S{\Sigma}
\def\ta{\tau}
\def\x{\chi}
\def\o{\omega}\def\O{\Omega}
\def\lagr{{\cal L}}
\def\cd{{\cal D}}
\def\k{\kappa}
\def\be{\begin{equation}}
\def\ee{\end{equation}}
\def\tz{\tilde z}
\def\tF{\tilde F}
\def\ri {\rightarrow}
\def\cf{{\cal F}}
\def\pa {\partial}
\begin{flushright}
FTUAM-96/24 \\
hep-th/9610032 \\ 
\today
\end{flushright}
\vspace{-1.2cm}

\begin{center}
{\large\bf IS THE STRING COUPLING CONSTANT INVARIANT UNDER 
T-DUALITY?}\footnote{Contribution  to the ``Second Meeting on Constrained
Dynamics and Quantum Gravity'', Santa Margherita Ligure, September 17-21,
1996 (E.A.) 
and to the
``XI International Workshop
on High Energy Physics and Quantum Field Theory''(QFTHEP'96; St. Petersburg, 
September 12-18, 1996)(Y.K.)}\\
\vskip .9 cm

\vskip .9 cm
{\bf Enrique Alvarez
\footnote{  
E-mail: enrial@daniel.ft.uam.es} and 
Yuri Kubyshin\footnote{On leave of absence from the Institute of Nuclear
Physics, Moscow State University, 119899 Moscow, Russia} \footnote{E-mail: 
kubyshin@delta.ft.uam.es}}.
 \vskip 0.05cm
Departamento de F\'{\i}sica Te\'orica, C-XI \\ 
Universidad Aut\'onoma de Madrid \\
28049 Madrid, Spain
\end{center}
\vskip .6 cm
\centerline{\bf ABSTRACT}
\vspace{-0.7cm}
\begin{quote}
\ \ \ \ \ It is well known that under  T-duality the sigma model 
dilaton (which is normally thought to be related to the 
string coupling constant through the simple formula 
$\kappa = \exp <\phi >$), undergoes an additive shift.
On the other hand, Kugo and Zwiebach, using a simplified form 
of string field theory, claim that the string coupling 
constant does not change under the T-duality.
Obviously, what seems to happen is that two different coupling constants,
associated to different dilatons, are used.
In this contribution we shall try to clarify this, and related issues.
\end{quote}

\normalsize
\newpage

\section{Introduction }

It is a well-known fact \cite{BU}, \cite{AO} in the theory of closed
bosonic strings propagating on a spacetime with a non-trivial group
of isometries, that an equivalent dual formulation exists.
The original and the dual models are related by
Buscher's formulas (which are not diffeomorphisms in general). For example, 
for an abelian isometry the dilaton undergoes a change
\be
\phi \rightarrow \phi^{\prime} = \phi - {1\over 2} \log k^2. 
\label{dilaton-tr} 
\ee
Here $k$ is the norm of the Killing vector associated to the isometry. 
In adapted coordinates, in which the Killing vector is 
$ {\pa\over \pa x^0}$, then  $k^2 = g_{00}$. 

There are basically three different (and complementary) 
ways of studying the physics of strings: 
in  first quantization , in the (target space) effective action 
formalism and in the 
string field theory (SFT) approach. Different dilatons exist  
in these three formulations  with corresponding coupling constants.
This means that we have several different couplings with 
slightly different meanings.
\par
In the first quantization 
the string coupling constant $\kappa$ is related to the dilaton through 
\be
\kappa = \exp <\phi >. 
\ee
This coupling  appears as the parameter of the expansion of the
partition function of the string over different world-sheet
topologies (in the closed case, this is equivalent to an expansion
over Riemann surfaces).  The relation above immediatly implies a
corresponding change in the (naive) string coupling constant under
T-duality:
\be
\kappa \rightarrow \kappa^{\prime} = \frac{\kappa}{ \sqrt{g_{00}}}
\ee
\par
There is further evidence for this transformation law, stemming from
the explicit computation by one of us \cite{AO} of the free energy
density of a string on a circle at arbitrary genus, an exact result.

In the SFT aproach, as we discuss below, the coupling $\lambda$
enters in a completely different way, as the coefficient of the
interaction term.  It was argued in Ref. \cite{KZ} that T-duality is
a symmetry of SFT and, as such, does not change the value of the
string coupling constant $\lambda$ (which, as has been repeatedly
emphasized by the authors   of
Ref. \cite{KZ}, must be, in principle, an observable). Although 
a simplified version of SFT (HIKKO
theory) is used, it is supposed to be powerful enough to tackle the
problem under consideration.

In sections 2-4 we  review the existing definitions of the dilaton in
various formulations of the string theory and discuss the coupling
constants associated to them as well as their mutual relationships.
We consider for simplicity closed bosonic strings in the critical
dimension $n \quad = \quad 26$ with $d$ spatial dimensions
compactified to a torus $T^{d}$ with common radius $R$, so that there
are $n - d$ non-compact dimensions.  We study in some detail the
change of dilaton under T-duality separately in each formulation, and
reach our conclusions in section 5.  A conjecture is made, in
particular, on the relationship between the ghost-dilaton field in
SFT and the sigma-model dilaton (based on previous results in the
linearized approximation).  It is not clear to us, however, how this
conjecture could be compatible with the implications of the
ghost-dilaton theorem in SFT.

We denote, as usual, the intrinsic string tension dimensionful
parameter by $\alpha'$ when neccessary. The couplings that we are
going to discuss are associated to the  vacuum expectation
values of the dilaton.  In that sense, they are characteristics of
the background and not another fundamental parameters of the string
theory.

\section{ First quantized bosonic string}

In the standard worldsheet formulation of the free closed string
moving in the $26$-dimensional target space in a background
characterized by the metric $G_{\m\n} (X)$, the antisymmetric tensor
$B_{\m\n}(X)$ and the (sigma-model or {\it curvature})- dilaton
$\Phi_{\s}(X)$ the action is given by \cite{GSW}
\begin{eqnarray}
S & = & \frac{1}{4\pi\alpha'} \int d^{2} \s \left[ \sqrt{\gamma}
\gamma^{\alpha \beta} G_{\mu \nu} \partial _{\alpha} X^{\mu}
\partial_{\beta} X^{\nu} + i \epsilon^{\alpha \beta} 
B_{\mu \nu} \partial _{\alpha} X^{\mu} \partial_{\beta} X^{\nu} 
\right. \nonumber \\
  & + & \left. \alpha' \sqrt{\gamma} \Phi_{\s} R^{(2)}
\right]. \label{action-ws}
\end{eqnarray}
In what follows we take $2\alpha' = 1$.
We consider the case when the background space-time has $d$ spatial
dimensions compactified to a torus of common radius $R$ and $ 26 - d$
non-compact dimensions, i.e. it is of the form $M^{26-d} \times
T^{d}$.

For  further discussion it is important to know the dependence of the
partition function $Z(R)$ on the radius $R$. Let us give a sketch of
this calculation \cite{AO}, \cite{VV}.  The contribution $Z_{g}(R)$
of the Riemann surface of genus $g$ is
\be
Z_{g}(R) = \int_{{\cal F}_{g}} d\mu F_{g}(R;\tau) 
\Lambda(\tau, {\bar \tau}),
\ee
where the integral is taken over the moduli space of the surface of
genus $g$, $\tau$ is the period matrix of the world-sheet Riemann
surface and $\Lambda(\tau ,{\bar \tau})$ is some function of $\tau$
and $\bar{\tau}$ only, whose explicit form is not important for us.
$F_{g}(R;\tau)$ is given by the functional integral over the string
field $X^{\m}$ satisfying the conditions
\begin{eqnarray}
X^{a}(z+a_{i},\bar{z}+a_{i}) & = & X^{a}(z,\bar{z}) + 2\pi R m_{i}^{a}, 
\label{xmn1} \\
X^{a}(z+b_{i},\bar{z}+b_{i}) & = & X^{a}(z,\bar{z}) + 2\pi R n_{i}^{a},
\label{xmn2}
\end{eqnarray}
where $a_{i}$ and $b_{i}$ ($i=1,2, \ldots g$) are  cycles generating
the homology of the Riemann surface  $\Sigma_g$, $a=1,2,\ldots , d$
and $m_{i}$, $n_{i}$ are the winding numbers:
\be
F_{g}(R;\tau) = \int DX e^{-S'(X)}.   \label{fg}
\ee
The action $S'$ is given by Eq. (\ref{action-ws}) but without the
dilaton term. The integral in (\ref{fg}) can be calculated by summing
over the contributions of the instanton solutions $X^{M}_{mn}$,
satisfying Eqs.  (\ref{xmn1}) and (\ref{xmn2}), and then integrating
over the fluctuations $x$ around the instantons:
\be
F_{g}(R;\tau) = \sum_{m,n} e^{-S'(X_{mn})} \int dq  \int Dx e^{-S'(x)}
\ee
The integral over the fluctuations is independent of $R$.  There are
two points which are important for us.  First, we indicate explicitly
the integration over the zero mode $q$ of the instanton solutions.
It appears because in each sector $(m,n)$ we have a continuous family
of  instanton solutions $X^{\m}_{mn} + q^{\m}$ with the same action
$S^{\prime}(X_{mn})$ parametrized by the 26-vector $q^{\m}$.
Integration over this zero mode gives the volume factor
\be
\int dq = V_{D} (2\pi R)^{d}.    \label{volume}
\ee
The appearance of this factor is in accordance with the calculations
in \cite{AW}.  We would like to stress here that the corresponding
factor in \cite{VV} and \cite{GRV} differs from the one in
(\ref{volume}), and is physically inequivalent to it.

Second, summation over $m$ and $n$ gives
\be
\sum_{mn} e^{-S'(X_{mn})} = \left( \det \; Im \; \tau \right)^{d/2}
\left(\frac{1}{( R\sqrt{2})^{g}} 
\sum_{mn} \exp f\left(\sqrt{2}Rm,\frac{n}{\sqrt{2}R} \right)\right)^d, 
\ee
where $f$ is a given symmetric function.  Notice that the sum in the
r.h.s. is obviously invariant under the $R \rightarrow 1/(2R)$
transformation. Tracing the dependence on the radius $R$ we obtain
that
\be
Z_{g}(R) = R^{d(1-g)} \tilde{Z}_{g}(R),     \label{Zg}
\ee
where $\tilde{Z}_{g}(R)$ is invariant under the T-duality
transformation $R \rightarrow 1/(2R)$. From the formula above it
follows that
\be
Z_{g}\left( \frac{1}{2R} \right) = (2R^{2})^{d(g-1)} Z_{g}(R).  
\label{Zg-transf}
\ee

If we add the dilaton term to the action (see Eq. (\ref{action-ws})),
the complete partition function for this background is given by the
sum
\be
  Z(\Phi_{\s}, R) \equiv \sum_{g=1}^{\infty} e^{2(g-1) \Phi_{\s}}
Z_{g}(R), \label{Z-sum}
\ee
where the exponent appears from the last term in the action
(\ref{action-ws}) with a constant dilaton background $\Phi_{\s}$
(recall that $\chi = (1/4\pi) \int d^{2} \s
\sqrt{\gamma} R^{(2)}$ is the Euler characteristic of the
two-dimensional manifold which for a compact Riemann surface of genus
$g$ is equal to $\chi = 2(1-g)$). Because of Eq. (\ref{Zg-transf})
the requirement that the T-duality is a symmetry of the string theory
implies that
\be
Z\left(\Phi_{\s} - \frac{d}{2} \log (2 R^{2}), \frac{1}{R} \right) 
= Z(\Phi_{\s},R). 
\label{Z-transf}
\ee
As it is known, all correlators can be obtained from the partition
function, and thus automatically transform correctly under the
T-duality transformation.

In this formulation the dimensionless string coupling constant is
defined through the vacuum expectation value of the background field
$\Phi_{\sigma}$ , that is:
\be
 \kappa = e^{<\Phi_{\s}>}.  \label{kappa}
\ee
Then the relation (\ref{Z-transf}) immediately leads to the
advertised change of the coupling constant, associated to a given
background, under the T-duality transformation:
\be
 \kappa \rightarrow \kappa' = \frac{\kappa }{(2R^{2})^{d/2}}.  
\label{kappa-transf}
\ee

Let us also comment here that the change of the dilaton
\be
\Phi_{\sigma} \rightarrow 
\Phi_{\sigma} - d\log (2R^{2}) /2,     \label{phi-transf}
\ee
which appears in Eq. (\ref{Z-transf}) as a consequence of the
T-duality invariance of the full partition function, coincides with
the shift of the dilaton (\ref{dilaton-tr}) due to the change of the
measure under the T-duality transformation in Buscher's construction,
see \cite{AAGL2}.
\footnote{At a fundamental level T-duality is 
essentially a canonical transformation, with the generating function
\be
F = {1\over 2} \oint d{\tilde \theta}\wedge d\theta
\ee
where the two Killing vectors in the original and the dual models are
$k={\partial\over\partial\theta}$, and ${\tilde k} =
{\partial\over\partial{\tilde\theta}}$  \cite{AAGL2},\cite{AAGL1},
and from this point of view the dilaton transformation
(\ref{dilaton-tr}) is rooted in the integration over momenta to
recover the lagrangian formulation.}

Given the fact that the determinant of the spacetime metric is
related to the one of its dual by the formula
\be
\sqrt{G} = k^2 \sqrt{G'}
\ee
there is an obvious combination which is duality invariant, namely:
\be
\phi_{inv} \equiv \Phi_{\s} -{1\over 4} \log G   \label{phi-inv}
\ee
\par
We would like to mention that the formulas (\ref{Z-sum}),
(\ref{Z-transf}) and (\ref{kappa-transf}) were also verified in a
discretization of the toroidally compactified string model in Ref.
\cite{AB}.

We conclude this section with discussion of the dilaton states.  It
appears that there are (at least) three states which qualify for the
name 'dilaton' in string  theory (\ref{action-ws}) 
 (see, for example,
\cite{BZ}, \cite{polchi}).  In particular, zero-momentum physical 
states correspond
to
\begin{enumerate}
\item the {\it ghost dilaton}, defined by the operator 
\be
D_g = {1\over 2} (c \pa^2 c - {\bar c} {\bar \pa}^2 {\bar c})
\label{ghost}
\ee
and 
\item the {\it matter dilaton}, defined by:
\be
\eta_{\mu\nu} {\mathcal{D}}^{\mu\nu} \equiv \eta_{\mu\nu} 
c{\bar c} \pa X^{\mu} {\bar \pa}X^{\nu}  \label{matter}
\ee
The relevant linear combination of the states above is 
\item the {\it zero momentum dilaton}, defined through:
\be
{\mathcal{D}} \equiv \eta_{\mu\nu} {\mathcal{D}}^{\mu\nu} + 
D_g  \label{zero}
\ee
\end{enumerate}
This field transforms as a scalar under spacetime diffeomorphisms. 
Another relevant combination is 
the {\it Graviton Trace}, defined by:
\be
{\mathcal{G}} \equiv \eta_{\mu\nu} {\mathcal{D}}^{\mu\nu} + 
{n\over 2}  D_g       \label{trace}
\ee
(Here $c$ and $\bar{c}$ are ghost fields, as usual).

\section{Low energy effective theory}

The effective action, describing the interactions of the massless
modes of the bosonic closed string in the lowest order in $\alpha'$
is equal to (see \cite{GSW}):
\be
S \equiv - {1 \over 2 \kappa_{26}^2} \int d^{n} x \sqrt{G} 
e^{-2\Phi} (R^{(G)} + 4 \pa_{\mu}
\Phi \pa^{\mu}\Phi - {3\over 4}H_{\mu\nu\rho}H^{\mu\nu\rho}) 
\label{action-s1}
\ee 
In the formula above the kinetic term for the gravitational field,
described by the ({\it string}) metric $G_{\mu\nu}$, is not
normalized in the standard way because of the exponential factor of
the dilaton in front of it; this can be unraveled by the use of the
{\it Einstein} metric, conformally related to the string metric as
follows:
\be
g_{\mu\nu} \equiv e^{- {4 \Phi\over n-2}} G_{\mu\nu}
\ee
In terms of the Einstein metric the effective action reads:
\be
S \equiv -{1\over 2\kappa_{26}^2} \int d^{n} x \sqrt{g} (R^{(g)} - 
{4\over n-2} \pa_{\mu}\Phi \pa^{\mu}\Phi - {3\over 4} e^{-{8 
\Phi\over n-2}} 
H_{\mu\nu\rho}H^{\mu\nu\rho})   \label{action-s2}
\ee
This ``target space dilaton'' (which is called also sometimes the
Fradkin-Tseytlin dilaton \cite{FZ}), represents the same $\s$-model
dilaton introduced in the previous section.  It is important to keep
in mind that, first, it is a scalar and, second, it transforms
according to (\ref{phi-transf}).

The coupling $\kappa_{26}$, which appears in front of the action
plays the role of the gravitational constant in 26-dimensional
space-time.  As it is known there is no actually any new dimensional
constant (apart from $\alpha'$). Indeed, similar to the first
quantized formulation, $\kappa_{26}$ can be always absorbed into the
vacuum expectation value $\Phi_{0}\equiv <\Phi >$ of the dilaton.
Thus, we can consider that for a given background characterized by
$\Phi_{0}$ we have the gravitational coupling
\be 
     \kappa_{26} = \kappa^{(0)}\,_{26} e^{\Phi_{0}}.
\ee
($\kappa^{(0)}\,_{26} \sim (\a^{\prime})^6$).
In the case of a space-time with $d$ dimensions toroidally
compactified (with dilaton $\Phi$ not depending on these $d$ compact
dimensions) the transformation $R \rightarrow 1/(2R)$ in the metric
$G_{\m\n}$ with the corresponding change of the dilaton $\Phi
\rightarrow \Phi' = \Phi - d\log (2R^{2}) /2$ (see Eq. 
(\ref{phi-transf})) leaves the action (\ref{action-s1}) invariant. In
this we should assume that the coupling is $\kappa^{(0)}_{26}$ and
does not change. Another point of view is to consider that in Eq.
(\ref{action-s1}) $\Phi = \phi$ is the quantum fluctuation of the
dilaton field and its vacuum expectation value is represented by the
gravitational coupling $\kappa_{26}$. Then the T-duality
transformation $R \rightarrow 1/(2R)$ does not affect the field
$\phi$, but on the  contrary the value of the coupling changes as
\be
\kappa_{26} \rightarrow \kappa_{26}' = 
\frac{\kappa_{26}}{(2R^{2})^{d/2}}, 
\label{kappa26-transf}
\ee
the same as for $\kappa$, Eq. (\ref{kappa-transf}). It is easy to see
from (\ref{action-s1}) that after the dimensional reduction to the
spacetime of $ 26 -d$ non-compact dimensions the effective coupling
constant becomes
\be
   \kappa_{26-d} = \frac{\kappa_{26}}{(2\pi R)^{d/2}}.   
\label{kappaD-def}
\ee
The dimensionally reduced theory in $M^{26-d}$ should not be affected
by the change of the radius of the compactified dimensions of the
initial 26-dimensional spacetime. Indeed, from the definition
(\ref{kappaD-def}) and the transformation law (\ref{kappa26-transf})
it follows that $\kappa_{26-d}$ is invariant.

\section{Dilaton in string field theory}

It is not unlikely that in the future {\it the} ``string field theory''
will be a quantum version of an eleven-dimensional M-theory, which is probably
not itself a theory of strings. But in the meantime, some impressive
achievements have been already made (see, for example,\cite{P1}- \cite{b9})
leading to a
consistent (non-polynomial) string field theory for closed strings.
\bigskip
In SFT  the free action is given by
\be
  S_{0} =  (( \Psi | {\cal K} | \Psi )),   \label{action-SFT}
\ee
where  ${\cal K} = Q + \tilde{Q}$ ,$Q$
and $\tilde{Q}$ are the BRST charges for the left-moving and
right-moving modes, and the scalar product $((.,.))$ is defined
in detail in \cite{Z}. The state $|\Psi>$, which describes the shift of
the original $\s$-model away from the empty flat spacetime, can be
written as
\be
| \Psi > = \left\{ ...+ h_{\mu \nu}(q) \alpha^{\mu}_{-1}
\bar{\alpha}^{\nu}_{-1} + \phi \left[ \bar{c}_{-1} b_{-1} +
c_{-1} \bar{b}_{-1} \right] \right\} c(0) \bar{c}(0) | 0 >, 
\label{psi-state}
\ee
We write only the part of $| \Psi >$ which is of interest for us, and
we use the standard notations for the coefficients of the mode
expansion for the string field and ghosts, now understood as
operators.  Notice that the states which appears in the expansion
(\ref{psi-state}), are the ghost dilaton, $|D_g> \equiv \phi D_{g}
|0>$ with $D_{g}$  given by (\ref{ghost}), and
$h_{\m\n}{\mathcal{D}}^{\m\n}$, where  ${\mathcal{D}}^{\m\n}$ can be
read off from (\ref{matter}).

The symmetric part $h_{\m\n}\,^{(s)}$ of the wave function $h_{\mu
\nu}$ can be shown to represent the graviton. This interpretation
appears in various ways. First of all it is known that physical
states are defined modulo a huge gauge invariance, generated by all
exact states in the BRST cohomology, that is, all states of the form:
\be
|\Psi>\quad + \quad (Q + {\bar Q})|\epsilon>
\ee
are physically equivalent. Thus, the gauge transformations, which
also leave the action (\ref{action-SFT}) invariant, have the form
\be
   \delta |\Psi > = (Q + {\bar Q})|\epsilon>.    \label{gauge-transf}
\ee
If we now choose
\be
|\epsilon> = \epsilon_{\mu}(\pa X^{\mu} c(0) + {\bar \pa}X^{\mu} 
{\bar c}(0)) | 0 >   \label{epsilon}
\ee
we generate a gauge transformation on the ``graviton'' state
\be
|grav> = h_{\mu\nu}\,^{(s)}{\mathcal{D}}^{\m\n} | 0 >
\ee
of the following form:
\be
\delta h_{\mu\nu} = \pa_{\mu}\epsilon_{\nu} + \pa_{\nu}\epsilon_{\mu} 
\label{h-transf}
\ee
and in the ``ghost-dilaton'' state,\, $|D_g> \equiv \phi D_{g} |0>$,
\be
\delta \phi =\partial_{\mu} \epsilon^{\mu}. 
\ee
These formulas show that the gauge transformation
(\ref{gauge-transf}), (\ref{epsilon}) lead to a general covariant
transformation and $\phi$ is not a scalar under this transformation.
If we make the field redefinitions
\begin{eqnarray}
  h_{\mu \nu}\,^{(s)} & = & \hat{h}_{\mu \nu}\,^{(s)} 
                          + \eta_{\mu \nu} \hat{\phi}, 
  \nonumber \\
  \phi & = & {\hat\phi} + \frac{1}{2} \hat{h}^{(s)}\,^{\mu}_{\mu}, 
\end{eqnarray}
then it is easy to check that the graviton field
$\hat{h}^{(s)}\,_{\mu \nu}$ has the same transformation law as
(\ref{h-transf}) and $\hat{\phi}$ is a scalar.  More detailed
considerations suggest that at least to the linear order the relation
between the fields $g_{\mu \nu}$ and $\Phi$ in (\ref{action-s2}) and
the string field (\ref{psi-state}) is given by
\begin{eqnarray}
g_{\mu \nu} & = & \eta_{\mu \nu} + \hat{h}_{\mu \nu}\,^{(s)} +
\ldots, \nonumber \\
\Phi - \Phi_{0} & = &\frac{n-2}{4} \hat{\phi} + \ldots = 
 -\frac{1}{2} \left( \phi - \frac{1}{2} h^{(s)}\,_{\m}^{\m} + 
\ldots \right)
 \label{relations}
\end{eqnarray}
where $\Phi_{0}$ is the constant background value of the dilaton.
These relations are also verified by the properties of the fields
under the $O(d) \times O(d)$ transformations \cite{Sen}.

In terms of these new fields the state (\ref{psi-state}) can be
written as
\be
|\psi> = \left\{ \ldots + ( \hat{ h}_{\m\n}\,^{(s)} - 
(1/n) \eta_{\m\n}
\hat{h}^{(s)}\,_{\rho}\,^{\rho})({\mathcal{D}}^{\m\n} - 
(1/n) \eta^{\m\n}
{\mathcal{D}}_{\rho}\,^{\rho}) + 
(1/n) \hat{h}^{(s)}\,_{\rho}^{\rho}
{\mathcal{G}} + \hat{\phi} {\mathcal{D}} \right\} |0>
\ee
with $\mathcal{D}$ and $\mathcal{G}$ are given by (\ref{zero}) and
(\ref{trace}).

The formulas above show that the ghost-dilaton field $\phi$ fails to
be invariant under (linearized) diffeomorphisms. On the other hand,
the form of the variation inmediatly conveys the fact that the zero
momentum dilaton
\be
|{\mathcal{D}}> \equiv \hat{\phi}{\mathcal{D}}|0>
\ee
is really invariant, $\d \hat{\phi}\, = \, 0$.  Further relations
between different formulations of the theory of strings come from the
calculation of the action (\ref{action-SFT}).  It was shown that this
action reproduces the part of the effective action (\ref{action-s2}),
containing the graviton and the dilaton, in the quadratic
approximation provided the relations (\ref{relations}) are taken into
account \cite{S1}, \cite{SZ} (see also \cite{P1}). 
Also in Ref. \cite{CLNY} it was demonstrated that the equation of
motion $ (Q + \tilde{Q})| \Psi > = 0$ in string field theory
coincides with the standard 1-loop beta-functions for a bosonic
string in background metric and dilaton fields with the same
identifications.

All this suggests that the $\s$-model dilaton and the dilaton field
in (\ref{action-s1})-(\ref{action-s2}) should be 
identified with the zero-momentum dilaton in
string field theory (in fact, with ${n-2\over 4}
\hat{ \phi}$), whereas the field $\hat{\phi}_{inv}$, Eq.
(\ref{phi-inv}), should be identified with the ghost dilaton,
$-{1\over 2}\phi$. (This agrees with Polchinski in \cite{polchi}).

Let us discuss now the T-duality transformation. According to the
idea of the string field theory approach the general formalism is
considered to be independent of the background \cite{KZ},\cite{bi}.
 As in usual
quantum field theory, the background appears when a general field
$|\Psi>$ is presented as
\be
|\Psi> = |\Psi'>\quad  + \quad |S>, \label{shift}
\ee
where $|S>$ describes the classical background and $|\Psi'>$
represents quantum fluctuations.  One goes from one background to
another by making a shift of the field of this type.  Now, the claim
of Ref. \cite{KZ} is that $O(d,d,Z)$ - transformations of the
background in general and T-duality transformation in particular do
not move the field $|\Psi>$ in the direction of the ghost dilaton,
specified by the states proportional to $\phi$ in Eq.
(\ref{psi-state}). This is in accordance with the T-duality
transformations in the $\s$-model formulation and the relation
between the ghost dilaton of the string field theory and $\phi_{inv}$
conjectured above.

Up to now we do not have any coupling constant in string field
theory. In this approach it is introduced, mimicking ordinary quantum
field theory,  through the definition of the interaction. There are
quite a few versions of the theory with interaction, starting from
the first covariant formulation by Siegel
\cite{S1}, \cite{S2}. It is worth mentioning, in particular, 
the one proposed by Witten \cite{Witten} for
open strings, which is based on differential geometry constructions.
A towering achievement has been the consistent formulation of a 
non-polynomial bosonic string field theory \cite{SFT-polynom}. 
A particularly simple theory (stemming from Siegel's picture)
 usually denoted  as $\alpha = p^{+}$ HIKKO
theory, \cite{HIKKO} has been argued by Kugo and Zwiebach
to serve the purpose of discussion of the T-duality. 

In the $\alpha = p^{+}$ HIKKO theory the interaction term
schematically is given by
\be
S_{int} = \frac{\lambda}{3} < V | |\Psi> |\Psi> |\Psi>,   
\label{inter-SFT}
\ee
where $<V|$ is the 3-vertex and we emphasize that the string coupling constant
$\lambda$ is defined through the three-point function (many
technical details of the construction are omitted). In the functional
integral a general shift of the type (\ref{shift}) followed by a
change of the field, made in order to maintain the invariance of the
kinetic term (\ref{action-SFT}), leads to a change of the coupling
constant $\lambda$.  Moreover, there are general statements, known as
ghost-dilaton theorems telling that the ghost dilaton is the only
BRST-physical state changing the string field coupling \cite{BZ},
\cite{BerZ}, \cite{RZ}.

In particular, if for the non-polynomial bosonic string the shift $|S>$
in (\ref{shift}) is along the ghost-dilaton: 
\be
\d |\Psi> \equiv |S> = \frac{\epsilon}{\lambda}|D_{g}>, 
\label{psi-shift}
\ee
then the corresponding change of the coupling constant is 
\be
\delta \lambda = \epsilon \lambda   \label{lambda-shift}
\ee
(\cite{BerZ}). The situation remains essentially the same for the 
$\alpha = p^{+}$ HIKKO theory \cite{KZ}.

The key point of the article \cite{KZ} is that the shift which
corresponds to the T-duality transformation of the background in
$\alpha = p^{+}$ HIKKO, does not involve states which change the
coupling constant (in particular it does not involve ghost dilaton
states, as it was said before).

With arguments, which still remain unclear to us, it was argued in
\cite{KZ} that the formula for the contribution $Z^{SFT}_{g}$ of a
string diagram with $g$ loops and $V$ vertices is given by
\be
Z^{SFT}_{g} = \lambda^{V} (\sqrt{G})^{g} \int_{x: \; \mbox{fixed}} 
{\cal D} X e^{-S}. \label{Zg-SFT}
\ee

The functional integral corresponds to the $Z_{g}(R)$ but with the
zero mode integration being carried out. For the interaction of the
type (\ref{inter-SFT}) (and only for this type, in fact) $V=2(g-1)$
for vacuum diagrams, and we have the correct power of $\lambda$,  in
agreement with Eqs. (\ref{Z-sum}), (\ref{kappa}).

\section{Conclusions}

As we have discussed at some length the simplest combination out of
the metric and the ghost dilaton which transforms as a scalar, is $
\phi - {1\over 2} \eta^{\mu\nu}h_{\mu\nu}$, so that it is tempting to
conjecture that, at the full nonlinear level, there is a relationship
between the ghost-dilaton and the sigma model dilaton of the type
\footnote{We would like to stress that our rather primitive methods do not
allow us to consider the full spacetime dependence of the 
observables involved; {\it all} conjectured relationships must be understood
to be valid only up to T-invariant quantities}
\be
\Phi_{\sigma} = -{1\over 2}(\phi -{1\over 2} \log \, G). 
\label{conjecture}
\ee
This relationship has previously been also considered in Refs. 
\cite{KZ}, \cite{Sen}, and, although we have only been able to argue 
for it in the linear
approximation, it would reconcile string field theory with sigma
model results.\footnote{It is known that in the closed string field theory,
an infinite number of (genus dependent) contact terms are needed to avoid
overcounting of Riemann surfaces. Detailed non-trivial calculations are 
obviously neccessary to check whether all these terms conspire to give
our conjectured result. We are grateful to Kelly Stelle for stressing
this point}

\par
If the conjecture made in (\ref{conjecture}) is true, 
the coupling $\kappa$ is associated with the zero-momentum dilaton.
Provided the SFT coupling $\lambda$ comes uniquely from the ghost
dilaton, as it is suggested by the discussion of the ghost-dilaton theorem 
in the previous section,
\be
\lambda = \mu (\phi),    \label{lambda-phi}
\ee
where $\mu$ is some function, 
we would find a relation of the type
\be
  \lambda = \mu \left( -2 \log \frac{\kappa}{G^{1/4}} \right) = 
   \mu \left( - 2 \log \frac{\kappa}{R^{d/2}} \right).   \label{kappa-lambda}
\ee
This implies that $\lambda$ is indeed invariant under
T-duality (cf. Eq. (\ref{kappa-transf})). 
For Eq. (\ref{Zg-SFT}) to be equivalent to (\ref{Z-sum}) 
the function $\mu$ in Eq. (\ref{lambda-phi}) must be chosen as 
\be
       \lambda = \mu(\phi) = \exp \left( -\frac{1}{2} \phi \right) =
    \frac{\kappa}{R^{d/2}}, 
   \label{mu1}
\ee
which is quite an expected form for a relation between 
the dilaton and the corresponding coupling constant\footnote{Let 
us notice {\it in passing} that eq. (\ref{Z-sum}) has a
form compatible with Weinberg's general unitarity arguments
\cite{W}.}. The discussion in the main body of the paper then 
would suggest that
$\lambda$ is associated with $\phi_{inv}$ (cf. Eq. (\ref{phi-inv})),
and with $\kappa_{26-d}$ in (\ref{kappaD-def}). 
\par
There is, however, a fact which we do not understand, and which may
point towards some inconsistency: the transformation of the
ghost-dilaton $\delta \phi = \frac{\epsilon}{\lambda}$ under the
shift (\ref{psi-shift}), and the corresponding change of the coupling
constant (\ref{lambda-shift}), as given by the ghost-dilaton theorem,
imply that 
\be
 \lambda = \mu(\phi) = \frac{1}{a-\phi} = 
   \frac{1}{a + 2 \log \frac{\kappa}{R^{d/2}}},  \label{mu2}
\ee
where $a$ is a constant. 
The relation of this form is rather peculiar. In particular 
we have week coupling regime $\lambda \rightarrow 0$ in string field 
theory both when $\kappa/R^{d/2} \rightarrow + \infty$ and 
$\kappa/R^{d/2} \rightarrow 0$.   

Thus, there is an obvious incompatibility between Eq. (\ref{mu1}) and 
Eq. (\ref{mu2}). 
We feel that it would be interesting to work out further details and
,clarify more this issue. After all, it seems that many string dualities
are in a sense a consequence of T-duality in a different context
(such as compactifications of M-Theory).
It is then obviously of the utmost importance to gain a knowledge of
T-duality in general settings as precise as possible.

{\bf Acknowledgements.}
We owe many thanks first of all to Barton Zwiebach for his remarks and 
useful suggestions, 
 and
to Luis Alvarez-Gaum\'e for his usual insightful comments, 
as well as to Yolanda Lozano and M. A. V\'azquez-Mozo for a careful 
reading of the manuscript.  Most helpful have also been
the detailed comments made by Tim Morris. Our work has been
supported by AEN/93/0673 and by  grants SAB95-0224 of M.E.C. (Spain) and 
96-02-16413-a of the Russian Fund of Fundamental Research.

\end{document}